\documentclass[twocolumn,showkeys,superscriptaddress,preprintnumbers,amsmath,amssymb,prl]{revtex4-1}
\usepackage{graphicx}
\usepackage{dcolumn}
\usepackage{bm}
\usepackage{amsmath}
\begin{document}

\title{Self-organized magnetic particles to tune the mechanical behaviour of a granular system}

\author{Meredith Cox}
\affiliation{Department of Physics \& Center for Nonlinear and Complex Systems, Duke University, Durham, North Carolina 27708, USA}
\author{Dong Wang}
\affiliation{Department of Physics \& Center for Nonlinear and Complex Systems, Duke University, Durham, North Carolina 27708, USA}
\author{Jonathan Bar\'{e}s}
\affiliation{Department of Physics \& Center for Nonlinear and Complex Systems, Duke University, Durham, North Carolina 27708, USA}
\author{Robert P. Behringer}
\affiliation{Department of Physics \& Center for Nonlinear and Complex Systems, Duke University, Durham, North Carolina 27708, USA}


\begin{abstract}
Above a certain density a granular material jams. This property can be
controlled by either tuning a global property, such as the packing
fraction or by applying shear strain, or at the micro-scale by tuning
grain shape, inter-particle friction or externally controlled
organization. Here, we introduce a novel way to change a local
granular property by adding a weak anisotropic magnetic interaction
between particles. We measure the evolution of the pressure, $P$, and
coordination number, $Z$, for a packing of 2D photo-elastic disks,
subject to uniaxial compression. Some of the particles have embedded
cuboidal magnets. The strength of the magnetic interactions between
particles are too weak to have a strong direct effect on $P$ or $Z$
when the system is jammed. However, the magnetic interactions play an
important role in the evolution of latent force networks when systems
containing a large enough fraction of the particles with magnets are
driven through unjammed states. In this case, a statistically stable
network of magnetic chains self-organizes and overlaps with force
chains, strengthening the granular medium. We believe this property
can be used to reversibly control mechanical properties of granular
materials.
\end{abstract}

\date{\today}

\keywords{Granular materials, Magnetic particle, Self organization, Jamming transition, Smart fluid}

\pacs{83.60.-a 81.05.Rm 45.70.Cc 05.65.+b} 

\maketitle


Above a certain density granular materials jam \cite{liu_nat1998}. As
the medium transitions from unjammed to jammed, its behavior changes
from fluid-like to solid-like. This property, which is often a
drawback in a silo or in a funnel, has been used recently for
innovative architecture constructions \cite{dierichs_ad2012} and soft
robotic actuators
\cite{Brown_pnas2010,cheng_icra2012,steltz_irs2009}. In these cases,
system parameters are tuned to reversibly change a soft material into
a solid and strong one. There are at least two ways to reach this goal
\cite{jaeger_sm2015}: (i) macroscopically: one can act on the material
boundaries to change the system volume/density \cite{Brown_pnas2010},
or one can apply system-wide shear strain ; (ii) locally: one can
change grain shapes \cite{miskin_nat2013,athan_sm2014}, local
organization \cite{dierichs_ad2012}, or friction. These different
approaches can be used to stabilize or destabilize the granular
packing. Reversibly changing the system properties at the global scale
is routine, but changing properties at the grain scale is much more
challenging. We take a first step toward a novel manner to affect the
macroscopic behavior of a granular packing by changing grain-scale
properties. We do so by the addition of magnets to a controlled
fraction of the grains. The magnetic interaction strength at the local
scale is very weak compared to interaction forces that are typical of
jammed states. Despite the low magnetic forces involved, the fluidity
of the granular system below jamming, provides a mechanism for
self-organization of structures that is reminiscent of
magneto-rheological fluids \cite{weiss_jimss1994,hagenbuchle1997_ao,ukai2004_pre}.


\begin{figure}
\centering \includegraphics[width=0.95\columnwidth]{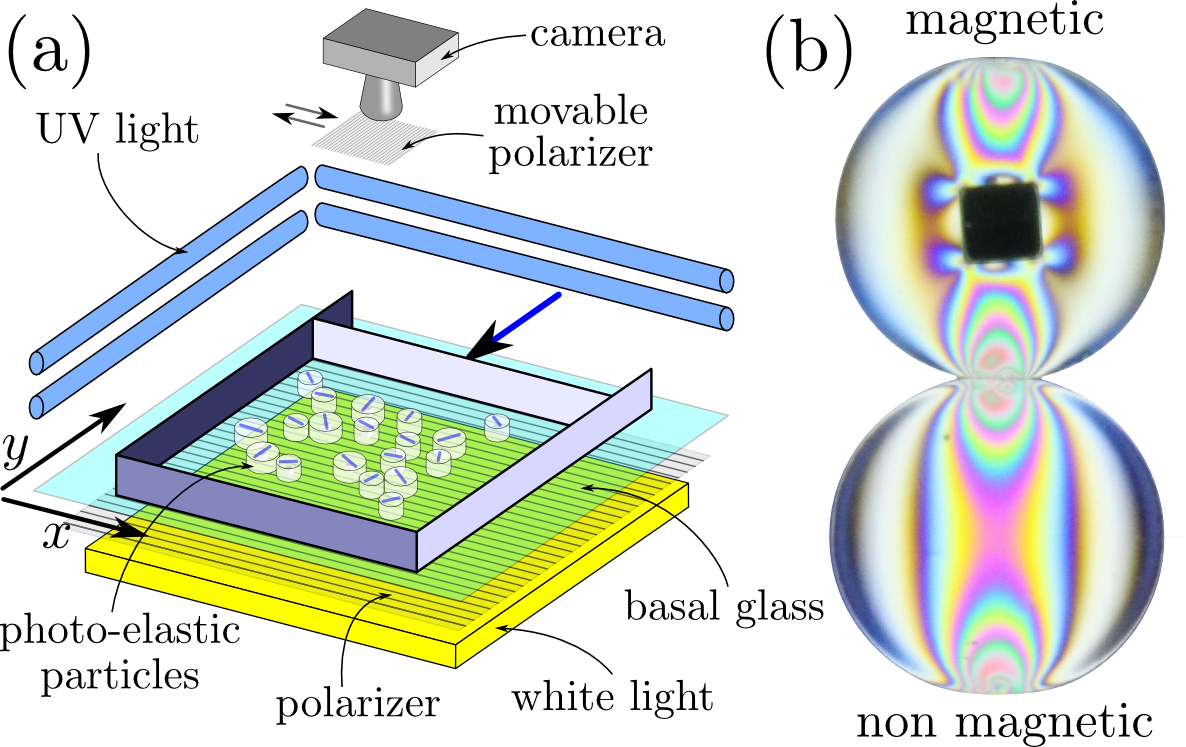}
\caption{(color online) (a): 3D schematic of the uniaxial compression
  experiment. A moving wall (blue arrow) compresses bidisperse
  magnetic photo-elastic particles by small steps. Each particle is
  marked on top by a bar, drawn in UV-fluorescent ink that gives the
  orientation of the embedded magnet. At each step the system is
  imaged in white light, (i) between crossed polarizers, (ii) without
  crossed polarizers, and (iii) in UV light. (b): top view with
  crossed polarisers showing magnetic and non-magnetic particles that
  have been compressed uniaxially. The presence of an embedded magnet
  modifies the inner mechanical and photoelastic response of a
  particle.}
\label{figExpe}
\end{figure}

The experiment consists of unixial compression (see
Fig.\ref{figExpe}-(a)) of a mixture of $645$ bidisperse ($60\%$ small,
$40\%$ large) photo-elastic disks with either diameter $12.7$ or
$15.9$~mm, and thickness $6$~mm. This ratio prevents crystalline
ordering. A $4 \times 4 \times 4$~mm$^3$ neodymium magnet was embedded
in the bottom of a fraction $R_m$ of the disks as in
Fig.\ref{figExpe}-(b), and a UV-fluorescent bar was drawn along the
attractive direction of the magnet on the top of each particle. The
particles were cast in a reverse-image chemically-resistant silicone
\cite{starmold} mold that was formed over a high resolution 3D printed
base. The disks were cast from shore stiffness $50$ liquid
polyurethane \cite{clearflex} and then cured. For some particles, a
magnet was centered at the bottom of the mold before pouring the
polyurethane; this process avoided residual stresses around the magnet
in the unstressed state. The friction coefficient between particles is
$0.62\pm0.07$.

\begin{figure}
\centering \includegraphics[width=1.05\columnwidth]{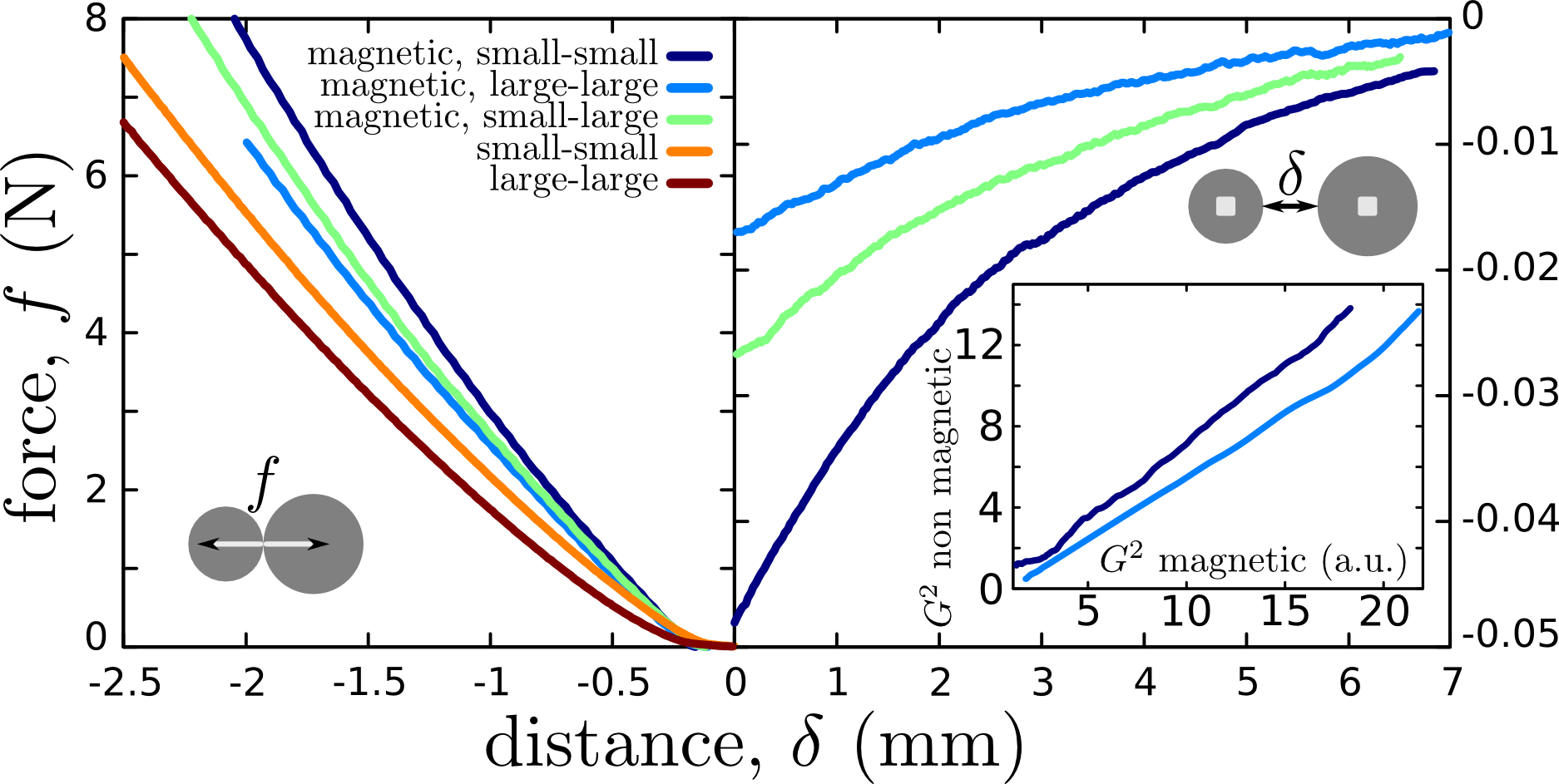}
\caption{(color online) Details of interactions between particles. For
  the different combination of particles
  (small/large/magnetic/non-magnetic) the interaction force is plotted
  as a function of distance $\delta$ or overlap ($\delta<0$) between
  particle boundaries. Force scale for positive $\delta$ (right panel)
  is not the same as for negative (left panel). Non-magnetic particles
  are slightly softer than magnetic once. For the latter, a maximum
  attractive force of $0.05$N is measured. Inset: for an equal
  compression force, $G^2$ (averaged over all pixels inside the
  photo-elastic materials of the particle) is plotted for magnetic
  $G_m^2$ and non-magnetic $G_{nm}^2$ particles, for small and large
  particles. A linear variation is found: $G_m^2=k \cdot G_{nm}^2$ with
  $k>1$.}
\label{figBehavPart}
\end{figure}

\begin{figure}
\centering \includegraphics[width=0.75\columnwidth]{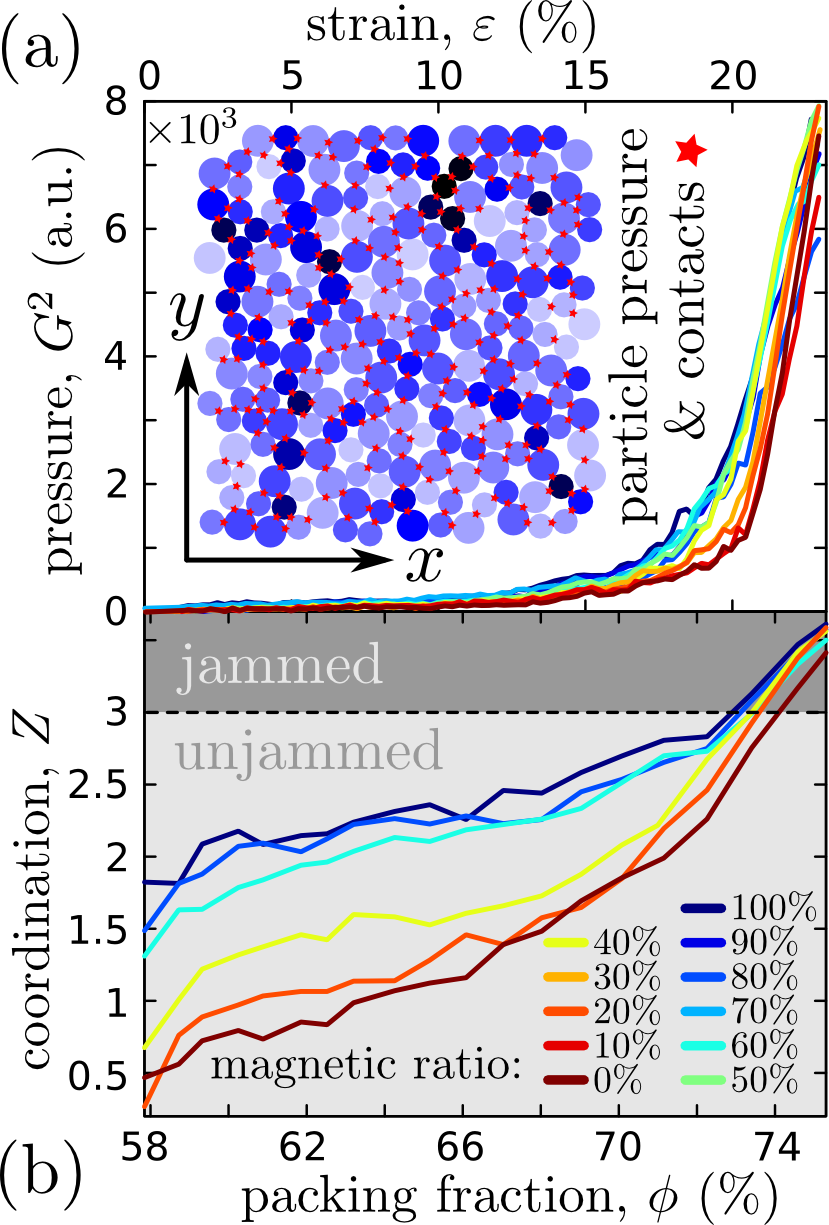}
\caption{(color online) Evolution of the pressure, measured as
  corrected $G^2$ (a) and of the coordination number, $Z$ (b) as a
  function of the applied strain $\varepsilon$ and of the packing
  fraction $\phi$ for different ratios of magnetic particles, from
  $0\%$ (only non-magnetic) to $100\%$ (only magnetic). The higher the
  percentage of magnetic particles, the stiffer the system and the
  faster pressure increases. The number of contact per grains is
  naturally higher for magnetic particles and the system jams
  faster. In (b) a dashed line for $Z=3$ marks the jamming
  transition. (a)-inset: part of the system for a $100\%$ magnetic
  particle experiment at $\varepsilon= 0.21$; darkness of the particle
  shows the pressure (from light--low pressure to dark--high
  pressure); red stars stands for contacts.}
\label{figPressCont}
\end{figure}

The disks were placed in a $44 \times 40$~cm$^2$ $(x,y)$ rectangular
cell, in a spatially random stress-free compact rectangle with the
long direction corresponding to the direction of
compression. Initially, there was a $\sim 3$cm gap between the
particles and the cell boundaries. The particles rested on a smooth
Plexiglas$^{\textsc{\textregistered}}$ sheet lubricated with talc,
yielding a particle-base friction coefficient of $0.36\pm0.04$. The
setup was illuminated from below by a circular polarized uniform white
light source, and from above by less intense UV lights. A $20$
megapixel SLR camera, mounted $2$~m above the particles, recorded
views with and without a crossed (with respect to the source) circular
polarizer (see Fig.\ref{figExpe}-(a)). After each $1$mm compression
step, the system was imaged a) without the second polarizer, b) with
the second crossed polarizer (see Fig.\ref{figExpe}-(b)), and c) with
the white light off and UV light on. The normal light pictures gave
the particle positions. The UV light pictures gave the magnetic
orientations.  The pressure inside each disk (see inset of
Fig.\ref{figPressCont}) was computed from the squared gradient of the
photo-elastic image intensity $G_m^2$ (for magnetic particles) and
$G_{nm}^2$ (for non-magnetic particles) following
\cite{howell_prl1999,geng_prl2001,Ren_prl2013,berhinger_jsm2014}. The
$P \sim G^2$ values were corrected using the inset graph of
Fig.\ref{figBehavPart} to account for differences in the different
particle types, as suggested by Fig.\ref{figExpe}-(b). That is, the
calibration between squared gradient and $P$ is different for
magnetic, $G_m$, and non-magnetic, $G_{nm}$, particles primarily
because the stiff magnet deforms the photo-elastic response. Contacts
between particle are also accurately detected (see inset of
Fig.\ref{figPressCont}) using positions and the photo-elastic
response, as in Majmudar et al. \cite{majmudar_prl2007}. The
sensitivity of the polyurethane permits us to observe a photo-elastic
response of the particles for applied forces as low as $0.015$N. This
is sufficient sensitivity to allow the detection of contacts due to
magnetic attraction of the particles.

The inter-particle interaction forces have been measured for pulling
and pushing different combinations of disks
(small/large/magnetic/non-magnetic) away/toward each other and
measuring the resulting force $f$ and edge-to-edge distance
between disks $\delta$ in a micro-strain analyser (TA Instruments RSA
III). Fig.\ref{figBehavPart} shows that during compression, magnetic
and small particles are stiffer. In every case the particles exhibit
Hertzian-like behavior and the magnetic attractive forces follow $f
\sim -e^{-0.35 \cdot \delta}$. Also, the highest attractive magnetic
forces ($0.05$N) are greater than forces due to basal friction ($\sim
0.005$N), and both are much lower than repulsive forces measured in
strong force chains ($\sim 2$N). This implies the magnetic forces
cannot play a direct role in the modification of the macroscopic
behavior of the granular material, once it is jammed.


However, the magnetic particles clearly influence the evolution of the
system and the medium response. Fig.\ref{figPressCont}-(a) shows the
evolution of the global pressure computed as the sum of $G^2$ over all
the disks when the granular system is compressed for different
magnetic particle ratios $R_m$, from $0\%$ (no magnetic particles) to
$100\%$ (only magnetic particles). Each curve represents an average
over three compression runs. Two different responses arise, one for
low $R_m$ and another for high $R_m$. The data for $R_m = 0\%$ is
characteristic of all the lower $R_m$ response curves, and $R_m=100\%$
is characteristic of the higher $R_m$ responses. In the first case,
$P$ stays around zero, and in response to strain, particles just
rearrange to relax the system like a liquid; $P$ then increases rather
rapidly above a critical-like $\varepsilon$. In the second case,
pressure increases more rapidly with $\varepsilon$, and the sharp
increase occurs at a lower $\varepsilon$ somewhat like a plastic
material.

The distinction between low and high $R_m$ is particularly dramatic
for the evolution of $Z$ vs. $\varepsilon$, or packing fraction
$\phi$, for different ratios of magnetic particles. As in
Fig.\ref{figPressCont}~(b), for $R_m<50\%$, and as previously observed
in \cite{majmudar_prl2007} for purely non-magnetic systems, $Z$ is
initially very low ($Z<1$) and increases over a modest range of
compressions to the jamming point $Z=3$ which corresponds to the sharp
increase in $P$. In contrast, when $R_m>50\%$ the coordination number
is always higher than in the non-magnetic case due to contacts caused
by magnetic attraction, which leads to $Z \simeq 2$. Then, when the
system is compressed, $Z$ increases roughly linearly until the rate of
increase accelerates after crossing the jamming point. As with $P$,
the responses are either nearly the same for all ($R_m<50\%$) or all
similar for all ($R_m>50\%$), with a rapid crossover for $R_m \simeq
50\%$. We believe this is due to a percolation transition for magnetic
chains we will describe later. The jamming packing fraction, $\phi_J$,
shifts from $\sim 0.728$ when the system is purely magnetic to $\sim
0.742$ when the system is completely nonmagnetic. Note that all the
jamming points are significantly lower than $0.84$, the isotropic
jamming point in $2D$, because uniaxial compression is anisotropic,
and friction between particles allows for shear jamming (note that
uniaxial compression is a mixed shear and isotropic compression
mode). There is also an effect from the friction between the particles
and the base. The basal friction is small, and the change of the
medium properties by replacing non-magnetic particles with magnetic
particles is clear. The transition between these two response types as
$R_m$ varies is remarkably sharp, which means that near the
transition, a dramatic change of the material response occurs with
only a very small change in the fraction of magnetic particles.


\begin{figure}
\centering \includegraphics[width=0.85\columnwidth]{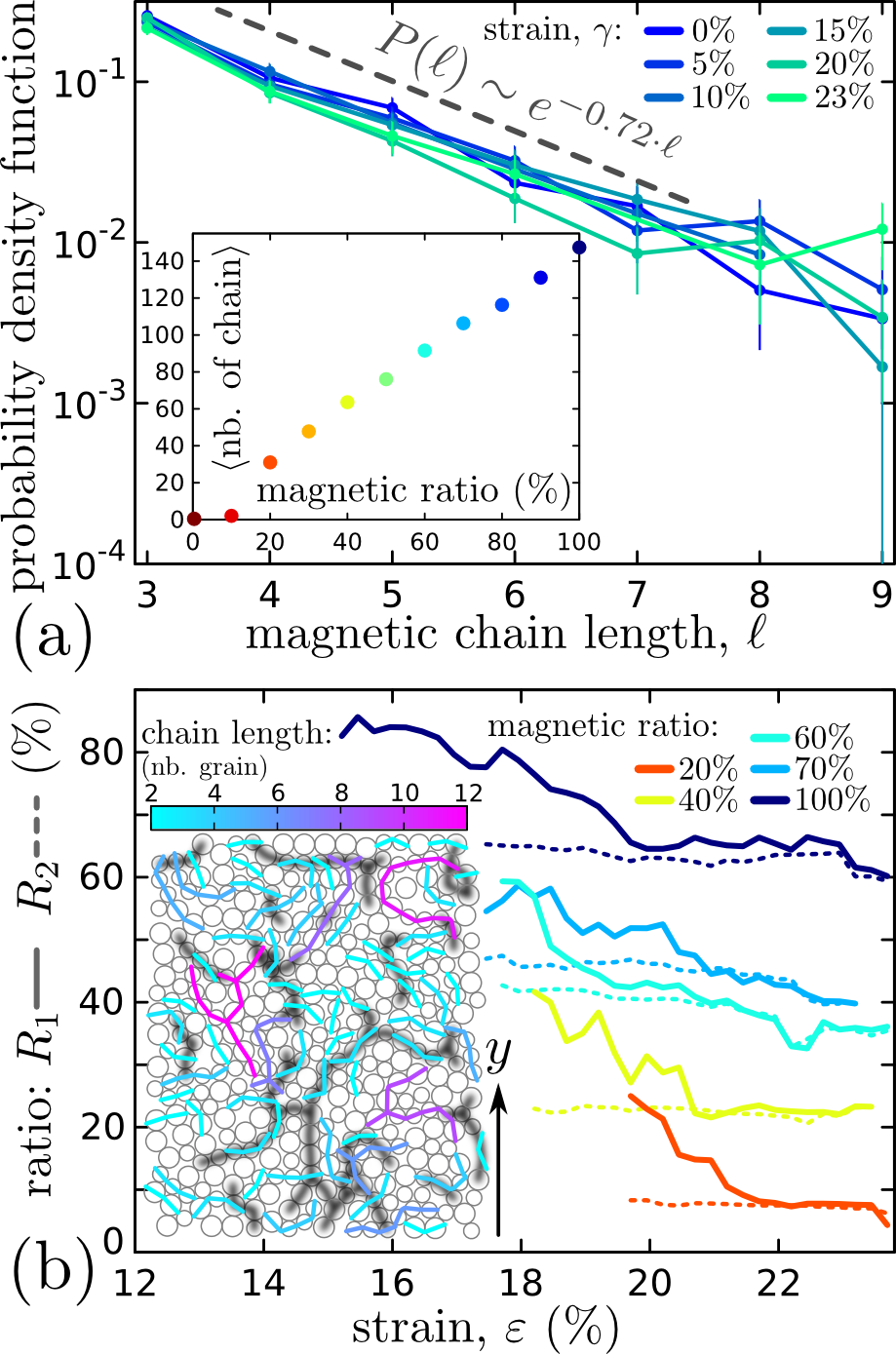}
\caption{(color online) (a): probability density function (PDF) of the
  length, $\ell$, of magnetic chains, averaged over $3$ $100\%$
  magnetic particle experiments measured at different steps of the
  compression. Even if chains break and merge, the PDF still follows
  an exponential law (grey dashed line) with coefficient $3.2 \pm
  0.15$. (a)-Inset: average number of magnetic chains for experiments
  with different percentage of magnetic particles. For each point, the
  average is done over all the compression steps for $3$ runs with the
  same $R_m$. The standard deviation is too small to be visible on the
  graph. (b): Evolution of the ratios $R_1$ (plain line, ratio of
  particles in a magnetic chain among particles in a force chain) and
  $R_2$ (dashed line, ratio of particles in a magnetic chain in the
  whole system) during the compression for different mixes of magnetic
  particles. Force chains mainly follow magnetic chains (see text for
  details). (b)-inset: part of the system for a $100\%$ magnetic
  particle experiment with $\varepsilon=16\%$, colored lines show
  magnetic chains (color scale for length) and dark lines shadow show
  force chains.}
\label{figChain}
\end{figure}

Although the macroscopic response when magnetic particles are
introduced changes significantly, as in Fig.\ref{figPressCont}-(a),
the magnetic forces are very small ($<0.05$N) compared to typical
strong repulsive forces ($\sim 2$N) in the jammed state. Hence, the
difference in $P$ and $Z$ induced by increasing $R_m$ cannot be
directly related to the weak magnetic forces. To understand the
changes in material response with $R_m$ we focus on the network
structures formed by the magnetic disks. As shown in the inset of
Fig.\ref{figChain}-(b), neighbour magnetic particles tend to align in
the same magnetic direction to attract each other, and hence to form
magnetic chains. A pair of particles in a magnetic chain are
identified as two nearby particles for which the distance between the
ends of their magnetic bars is less than $3$mm. These chains can
consist of up to $12$ particles, and can exhibit ramifications and
loops (see inset of Fig.\ref{figChain}-(b)). We define the length,
$\ell$, of a magnetic chain as the number of member particles, and we
consider only chains that have $\ell \ge 3$. Fig.\ref{figChain}-(a)
shows the evolution of the probability density function (PDF),
$P(\ell)$, for different compression steps for an $R_m = 100\%$
packing. The statistical distribution of $\ell$ is constant during the
compression even if chains break and merge during the process (see
video in supplementary material). $P(\ell)$ follows an exponential law
with coefficient $0.72\pm0.0115$:

\begin{equation}
P(\ell)=(2.0 \pm 0.013) e^{-0.72 \cdot \ell} 
\end{equation}

\noindent This statistical distribution remains the same within the
errorbars for all $R_m > 0$ for the present experiments (not shown
here).

This implies that whatever their density and the constraints inside
the system, magnetic particles self-organize to maintain a magnetic
chain network of the same statistical nature. Also, the average number
of magnetic chains is roughly constant during the compression process
but as shown in the inset of Fig.\ref{figChain}-(a) it increases
linearly with $R_m$.

The external load on a granular medium is not supported homogeneously
by all the grains of the system but along forces chains forming a
sparse network
\cite{radjai_prl1998,majmudar_prl2007,kondic_epl2012}. The force
network presented as a black shadow in the inset of
Fig.\ref{figChain}-(b) is formed by particles having a $G^2$ value
(averaged over all pixels inside the photo-elastic materials of the
particle) higher than a certain threshold ($8$a.u., $60$\% of the
maximum $G^2$ value in Fig.\ref{figChain}-(b)) and in contact with
other particles in the same situation. The threshold is far above the
pressure that could result from magnetic interactions.

There are several seemingly contradictory facts that need explanation:
1) the magnetic forces are too small to directly affect the system
response; 2) $P$ and $Z$ as functions of strain depend significantly
on $R_m$; 3) the magnetic particles self-organize into statistically
stationary networks. A reasonable way for the magnetic particles to
strongly affect the pressure, would be if they were associated with
the strong force networks. To test this hypothesis, we investigated
the effect of the magnetic networks on the force network
organization. Fig.\ref{figChain}~(b) shows that when force chains
begin to form, they grow preferentially along magnetic chains. We
define $R_1$ to be the ratio of the number of particles in magnetic
chains that are also in force chains to the number of particles in
force chains. We compare this ratio to $R_2$, defined as the ratio of
particles in magnetic chains to all particles in the total system:

\begin{equation}
	\begin{split}
		& R_1=\dfrac{\mathcal{N}(\text{particles in force \& magnetic chain})}{\mathcal{N}(\text{particles in force chains})} \\
		& R_2=\dfrac{\mathcal{N}(\text{particles in magnetic chain})}{\text{total number of particles}}
	\end{split}
\end{equation}

\noindent Here, $\mathcal{N}$ means particle number. If force chains form everywhere in the system without regard to magnetic chains, then those two ratios should be equal.

But as shown in Fig.\ref{figChain}-(b), initially, $R_1$ is much
higher than ratio of magnetic chain particles in the total system:
$R_1 \gg R_2$. Then the difference between both ratios vanishes when
the pressure increases, and the network becomes denser after the
jamming transition. This means that the first force chains of the
granular material, which are the ones that constitute the backbone of
the system and that will be the most loaded when the pressure
increases, form mainly along the magnetic chains, whatever the ratio
of magnetic particles.

Hence, the force chains that constitute the mechanical structure of a
granular system under loading is imposed by a magnetic chain structure
that itself self-organized and was statistically stationary during the
evolution of the packing. The nearly binary response of the granular
medium (like either the $R_m=100\%$ or $0\%$ responses) suggests that
for low $R_m$, the density of magnetic chains, even if force chains
form along them, is too low to be able to setup a strong structure and
hence dominate the macroscopic response of the material. The sharp
transition as $R_m$ is varied smoothly (around $R_m=50\%$) to a very
different response strongly suggests a percolation transition. This
property could be used to tune the rigidity of a granular material by
changing the magnetic nature of a small number of grains. This tuning
could be carried out by any process (e.g. temperature change, external
magnetic field change) that would change the magnetic state of a small
number of particles.

To conclude, we observe that even though the magnetic inter-particle
force for our particles is much lower than typical jammed-state
repulsive forces, the magnetic particles have an important impact on
the response to compaction of the granular material, provided the
fraction of magnetic particles is large enough. More precisely, a
sharp transition, that we believe can be mapped to a percolation
transition in the pressure/coordination-strain/packing fraction
curves, is observed around $R_m \simeq 50\%$. Also, self-organized magnetic
chains occur in the packing at a very early stage. Even though the
magnetic chains evolve during compression (breaking, merging) the
statistics of their length remain the same throughout the
process. This last point helps explain the first one, since we observe
that force chains that support the applied load are formed
preferentially along the magnetic chains, particularly for smaller
strains. When the magnetic chain fraction is too low, they cannot
impose a mechanical network that is strong enough to affect the
macroscopic behavior of the granular material. We believe this
property can be used in novel applications.


Acknowledgements: This work was supported by The NSF-Research Triangle
MRSEC, NSF grant DMR-1206351, The William M. Keck Foundation, and a
London post-doctoral fellowship for JB.

\bibliographystyle{unsrt}
\bibliography{biblio}

\end{document}